\documentclass[12pt]{article}

\makeatletter\@addtoreset{equation}{section}\makeatother

\newcommand{\beq}{\begin{equation}}
\newcommand{\eeq}{\end{equation}}
\newcommand{\ber}{\begin{eqnarray}}
\newcommand{\eer}{\end{eqnarray}}

\newcommand{\eeql}[1]{\label{#1}\eeq}

\newfont{\Bbb}{msbm10 scaled 1200}     
\newcommand{\mathbb}[1]{\mbox{\Bbb #1}}

\input epsf

\newcommand{\figin}[2]{
\begin{figure}[t]
\centerline{\hbox{\epsffile{#1.eps}}}
\centerline{\parbox{10cm}{\caption{#2\label{#1}}}}
\end{figure}}

\newcommand{\fig}[1]{Fig.~\ref{#1}.}
\newcommand{\figur}[2]{\fig{#1}\figin{#1}{#2}}

\newcommand{\bra}[1]{{\left< {#1} \right|}}
\newcommand{\ket}[1]{{\left| {#1} \right>}}

\newcommand{\cC}{{\cal C}}

\newcommand{\cQ}{{\cal Q}}

\newcommand{\half}{{1\over 2}}

\begin{document}

\begin{titlepage}

\begin{table}[t]
\begin{flushright}
   WIS/01/03/JAN-DPP\\
   hep-th/0301079
\end{flushright}
\end{table}

\begin{center}
{\LARGE
On Different Actions for the Vacuum of 
\\\vskip5mm
Bosonic String Field Theory}
\vspace{10mm}

\normalsize{
Nadav Drukker}

\vspace{5mm}

{\em Department of Particle Physics, \\
Weizmann Institute of Science, \\
Rehovot 76100 Israel}

\vspace{5mm}
{\tt drukker@weizmann.ac.il}

\end{center}

\vspace{10mm}

\begin{abstract}
We study a family of kinetic operators in string field theory describing 
the theory around the closed string vacuum. Those operators are based 
on the analytical classical solutions of Takahashi and Tanimoto and are 
analogous to the pure ghost action usually referred to as ``vacuum string 
field theory,'' but are much more general, and less singular than the 
pure ghost operator. The closed string vacuum is related to the D-brane 
vacuum by large, singular, gauge transformations or field redefinition, 
and all those different representations are related to each other by 
small gauge transformations. We try to clarify the nature of this 
singular gauge transformation. We also show that by choosing the Siegel 
gauge one recovers the propagator proposed in hep-th/0207266 that generates 
closed string surfaces.
\end{abstract}

\end{titlepage}

\section{Introduction}

The usefulness of string field theory \cite{Witten:1985cc}
as more than just a 
rewriting of first quantized open strings was demonstrated in recent 
years. There is little doubt now that string field theory includes 
classical solutions, or vacua, with any number of D-branes. The theory 
is defined around the D-brane vacuum and a lot of work has gone into 
finding the description around the lower vacuum, without the D-brane.

A classical solution describing this vacuum was found in successive 
numerical approximations using level truncation in the Siegel gauge 
based on \cite{Kostelecky:1989nt}
(for the most detailed calculation see \cite{Gaiotto:2002wy} and 
references therein). 
Another approach to the problem \cite{Rastelli:2000hv}
was to guess the form of the action 
around the new vacuum. Since the action is cubic, by expanding it 
around the classical solution $\Phi_0$ one finds the same form 
of the action but instead of the BRST charge $Q_B$ the kinetic operator
is given by
\beq
\cQ\Psi=Q_B\Psi+\Phi_0\star\Psi-(-1)^\Psi\Psi\star\Phi_0\,.
\eeql{shift}
One then took the simplifying assumption that $\cQ$ is pure ghost, and 
found \cite{Hata:2001sq,Gaiotto:2001ji} that it should be the midpoint 
insertion of $c$, that is $c(\pi/2)$.

Using this action one could construct the classical solutions 
describing D-branes and calculate their actions \cite{Okawa:2002pd}. 

The situation is still not fully satisfactory. For one, it is of 
interest to find the analytical form of the 
solution starting from the open string action. In principle, 
if one identifies all the open string states living on the sliver one 
would have the full field redefinition between the two forms of the 
the action, and the analytical classical solution, but that seems 
very hard and tedious. The second objection is that the pure ghost kinetic 
term is very singular, so one might hope to find nicer expressions for the 
closed string vacuum.

This was in fact done in a very nice paper of Takahashi and Tanimoto 
\cite{Takahashi:2002ez}. They construct classical solutions of bosonic 
open string field theory by acting with certain combinations of the 
BRST current and the ghost 
on the identity state. They were able to show that those solutions are 
gauge transformations of the usual vacuum, but sometimes those gauge 
transformations are singular, and lead to non-trivial solutions.

They also used those solutions to find the kinetic operator around the 
new vacuum, as in (\ref{shift}). Again those are sometimes equivalent 
to the regular kinetic operator $Q_B$, but when the gauge transformation 
is singular, they are not. In particular, in one case it was shown in 
\cite{Kishimoto:2002xi} that this new kinetic operator has trivial 
cohomology at the right ghost number. This is the hallmark of the 
kinetic operator around the closed string vacuum, since we should not 
find any on-shell open strings in the spectrum.

In this paper we study those kinetic operators.

The classical solutions they found are based 
on the identity state, and are rather hard to control. For example the 
value of the classical action, which should be minus the tension of the 
D-25 brane, has not yet been calculated. But the kinetic operators are 
smooth and in many ways nicer than the pure ghost $c(\pi/2)$.

The next section is devoted to the basic properties of those kinetic 
operators. 
We present them as constructed in \cite{Takahashi:2002ez}, as a 
convolution of the BRST current with some function $f$ (plus a pure 
ghost term). There is such an operator $\cQ_f$ for any function $f$ 
on the interval $[0,\pi]$ which is equal to one at the midpoint and 
symmetric around it. We also review how these operators are related 
to the usual $Q_B$ by a field redefinition.

We then study some properties of those field redefinitions and under 
what conditions they are singular. The answer turn out to be very 
simple, that the function $f$ have a zero or a pole.

One example we point out is that by a singular limit on the function 
$f$, taking it to be zero everywhere except the midpoint, we can get 
the pure ghost operator of \cite{Hata:2001sq,Gaiotto:2001ji}.

In section~3 we study the gauge fixed action. By choosing the Siegel 
gauge one finds the gauge fixed kinetic operator which includes the 
convolution of the energy momentum tensor with the function $f$. This 
in fact fits beautifully with the ideas in \cite{Drukker:2002ct}, 
where a similar type of gauge-fixed kinetic operator was derived from 
totally different considerations. There it was argued that the kinetic 
operator will not be gauge equivalent to the usual one if $f$ has 
zeros on the boundaries of the world-sheet. And instead of open 
surfaces, the Feynman rules derived from this action will generate 
closed surfaces.

By starting with the exact gauge invariant form of the action we will 
fix some extra terms in the propagator, related to the ghosts, 
which were not determined in \cite{Drukker:2002ct}.

Another point, related to closed string vertices, is clarified in 
section~4. The open-closed vertices of Shapiro and Thorn are gauge 
invariant only when they commute with the kinetic operator. In the 
D-brane vacuum the commutator with $Q_B$ leads to the mass-shell 
condition for the closed string operators. The singular pure 
ghost operator does not give this constraint on the closed string 
spectrum, but we show here that all the $\cQ_f$ do indeed satisfy 
this condition and pick out only on-shell closed string vertices.

In section~5 we go back to the singular field redefinitions and try to 
elucidate how the zeros of $f$ lead to the non-trivial vacua. The field 
redefinition that puts the action back in the original form, with the 
BRST charge involves an operator built out of the ghost current and 
the logarithm of $f$. When multiplying fields with the star product 
those operators cancel each other for regular $f$. But for singular 
functions the operators are defined on different branches on the 
logarithm, so they cannot cancel each other completely.

We learn two things from this calculation. First, it proves that only 
the singularity structure of $f$ matters. Small gauge transformations 
can change the value of $f$, but adding and removing zeros are large 
gauge transformations. We conjecture that the number of zeros and their 
order is related to the number of D-branes, or the rank of the 
projector. Second, it allows us the write the action around the closed 
string vacuum in a new way. The new action has the usual kinetic 
operator $Q_B$, but a modified gluing rule in the star product and 
integration.

The last section is devoted to some discussion and more speculations.

\section{Kinetic operator around the vacuum}

\subsection{Review}

In \cite{Takahashi:2002ez} Takahashi and Tanimoto constructed a family 
of classical solutions to Witten's cubic string field theory. Then 
they proceeded to formulate the action for small fluctuations around their 
classical solutions. Since the construction is rather formal and 
technical, I will not repeat it here, but refer the reader the the 
original paper. The outcome of the calculation is an action like 
Witten's
\beq
S=\int \Phi\star \cQ_f \Phi +{2\over3}\Phi\star\Phi\star\Phi\,.
\eeq
Here $\Phi$ is the string field shifted by the classical solution 
$\Phi_0$, and the integration and star product are the regular gluing 
rules for string fields \cite{Witten:1985cc}. The only new ingredient 
is the kinetic operator $\cQ_f$ defined in terms of the BRST current 
$j_B$ and the ghost $c$ as
\ber
\cQ_f
&=&{1\over 2\pi}\int_0^\pi d\sigma 
\left[(j_B(\sigma)+\bar j_B(\sigma))f(\sigma)
-(c(\sigma)+\bar c(\sigma)){f'(\sigma)^2\over f(\sigma)}\right]
\nonumber\\*
&=&{1\over2\pi i}\oint dw \left[j_B(w)f(w)-c(w){f'(w)^2\over f(w)}\right]\,.
\eer
In the first line we wrote the integral over the coordinate $\sigma$ on 
the strip, and in the second line we represented the string field as half 
the unit disc in the upper half plane. The coordinates are related by 
$w=\exp i\sigma$, and we included the antiholomorphic piece by an image 
in the lower half plane.

The particular choice $f=1$ gives the usual kinetic operator 
$Q_B=1/(2\pi i)\oint dw j_B(w)$.

The fact that $\cQ_f$ is nilpotent and a derivation of the star algebra 
are direct consequences of its construction as a shift of the regular 
kinetic operator (\ref{shift}). But it is also not too hard to verify it 
directly. Using the OPEs
\ber
j_B(z)j_B(w)
&\sim&{-4\over (z-w)^3}c\partial c(w)-{2\over (z-w)^2}c\partial^2 c(w)
=-\partial_w\left({2\over (z-w)^2}c\partial c(w)\right)\,,
\nonumber\\*
j_B(z)c(w)&\sim&{1\over z-w}c\partial c(w)\,,
\eer
one immediately sees that the commutator of the term coming from the 
two currents cancels the cross term.

The other important property is that $\cQ_f$ be a derivation of the 
star algebra, that is 
$\cQ_f(A\star B)=\cQ_fA\star B+(-1)^AA\star\cQ_f B$. This can be 
expressed in terms of the three-string vertex $\bra{V_3}$ as
\beq
\bra{V_3}\left(\cQ_f^{(1)}+\cQ_f^{(2)}+\cQ_f^{(3)}\right)=0\,,
\eeq
where the superscript on $\cQ_f$ indicates which of the three strings 
it acts on. This equation is a consequences of the conservation laws 
derived in \cite{Rastelli:2000iu} for the $c$ ghost, and the analogous 
equalities for the BRST current. When considering the pure ghost 
operator \cite{Rastelli:2000hv} this imposed the constraint that only 
the combinations $\cC_n=c_n+(-1)^n c_{-n}$ of the modes of the ghost 
appear. A similar constraint exists for the modes of the BRST current. 
Those imply that the function $f(w)$ has to satisfy $f(-1/w)=f(w)$.

Using an identical argument one can show that $\cQ_f$ annihilates the 
identity state. This implies that the new action has the gauge 
invariance (at the linear level) of a shift by $\cQ_f\Lambda$ for any 
$\Lambda$.

Another condition on $f$ that comes out of their construction is that 
$f(i)=f(-i)=1$. We have not found an independent explanation of it.

\subsection{Singular field redefinitions}

As is pointed out in \cite{Takahashi:2002ez}, one can relate $\cQ_f$ 
to the usual $Q_B$ by the similarity transformation
\beq
\cQ_f=e^{q(h)} Q_B e^{-q(h)}\,,
\eeq
where $q(h)$ is an operator constructed out of the ghost current 
$j_{gh}=cb$ convoluted with $h=\log f$ as
\beq
q(h)={1\over 2\pi i}\oint dw j_{gh}(w) h(w)\,.
\eeq
The proof follows from repeated use of the commutators 
$[q(h),j_B(w)]=h(w)j_B(w)+2\partial(c(w)\partial f(w))$ 
and $[q(h),c(w)]=h(w)c(w)$.

In order to define $e^{q(h)}$ properly in the quantum theory they wrote 
it in normal ordered form. First they separate $q(h)$ into the sum of the 
zero mode and positive and negative modes as 
$q(h)=q_0(h)+q^{(+)}(h)+q^{(-)}(h)$, then one can define
\beq
e^{q(h)}=e^{\half\left[q^{(+)}(h),q^{(-)}(h)\right]}
e^{q_0(h)}e^{q^{(-)}(h)}e^{q^{(+)}(h)}\,.
\eeq
They noted that in some cases the normal ordering constant from the 
commutator of the positive and negative frequencies diverges. If we 
expand $h$ and $j_{gh}$ as
\beq
h(w)=\sum h_n w^{-n}\,,
\qquad
j_{gh}(w)=\sum q_n w^{-n-1}\,.
\eeq
then using $[q_n,q_m]=m\delta_{m+n}$ we see that the commutator is
\beq
\left[q^{(+)}(h),q^{(-)}(h)\right]
=\sum_{n=1}^{\infty}nh_{-n}h_n\,.
\eeql{com}

If the function $f$ vanishes on the unit circle, the function $h$ will 
have a logarithmic singularity, and the Laurent coefficients will 
behave as
\beq
h_{-n}\sim h_n\sim{1\over n}\,.
\eeq
This behavior will lead to a divergence in the normal ordering 
constant (\ref{com}), and $e^{q(h)}$ will be ill defined.

We therefore conclude that if $f$ has no singularities on the unit circle 
the action with $\cQ_f$ is a regular field redefinition of the usual 
action, so it describes the open strings on the D-25 brane. If 
$f$ has singular points this action could describe a different vacuum.

in particular for the function
\beq
f=-{1\over4}\left(w-{1\over w}\right)^2 = \sin^2\sigma\,,
\eeq
it was shown in \cite{Kishimoto:2002xi} that 
$\cQ_f$ has trivial cohomology at ghost number one. Therefore the action 
with this kinetic term is an appropriate candidate to describe the 
closed string vacuum. Any other function with the same zeros will 
yield the same results, since they will be related to each other by 
a regular field redefinition.

It is not hard to see the source of the problem. The action of 
$e^{q(h)}$ on the ghost $c$ is
\beq
e^{q(h)}c(w)e^{-q(h)} = f(w)c(w)\,,
\eeq
so if the function $f$ has zeros or poles on the unit circle it is a 
singular operator (acting on the anti-ghost $b$ will multiply it by 
$1/f$). We believe that the source of the singularity is the attempt to 
write an operator with a non-trivial kernel as an exponent. 
In Section~5 we try to make sense of this operator by deforming the 
contours used to define $q(h)$.

It is not surprising that a singular transformation changes the 
cohomology. If we write $Q_B=e^{-q(h)}\cQ_fe^{q(h)}$, and assume that 
$\cQ_f$ has trivial cohomology. The new physical states should come 
from the kernel of the transformation $e^{q(h)}$.

\subsection{Pure ghost $\cQ$}

One may wonder about the pure ghost kinetic term of 
\cite{Hata:2001sq,Gaiotto:2001ji}. How is this 
related to those operators. In fact it is a limit of such operators.

Consider a function $f$ that is close to zero along the entire unit 
circle, except for the midpoint ($i$ and $-i$). For example
\beq
f={\left(w^2-1\right)^2\over
\left(w^2-1\right)^2-a^2\left(w^2+1\right)^2}\,,
\eeq
with very large $a$. The integral of $f$ around the circle is very 
small (proportional to $1/a$), so there is no contribution to $\cQ_f$ 
from the BRST current $\oint dw j_B f$, but the derivative of $f$ 
diverges near the midpoint, so there is an infinite contribution from 
\beq
\cQ_f\sim\oint dw {(f')^2\over f}c
\sim a \left(c\left({\pi\over2}-{1\over a}\right)
+c\left({\pi\over2}+{1\over a}\right)\right)\,.
\eeq
In the limit this can be regarded as the pure ghost midpoint operator.

\section{Gauge fixed kinetic operator}

given this action one may try to check different properties of it. One 
possible check is to derive the Feynman rules, as was done 
for the pure ghost operator in \cite{Gaiotto:2001ji}. In particular 
we will compare the results with \cite{Drukker:2002ct}, where 
the general form of the gauge fixed action, and of the Feynman 
rules were proposed.

The Siegel gauge condition on a string field is $b_0\Phi=0$, where $b_0$ 
is the zero mode of the anti-ghost. This means that any field satisfying 
the gauge condition can be written as $b_0\Psi$ for some $\Psi$. Therefore 
the quadratic part of the action $\bra{\Psi}b_0 \cQ_f b_0\ket{\Psi}$
picks out the term in $\cQ_f$ proportional to $c_0$, the ghost zero 
mode. To calculate it we use the OPE
\beq
j_B(z) b(w) = {3\over(z-w)^3} + {j_{gh}(w)\over(z-w)^2}
+ {T(w)\over z-w} +\dots
\eeq
So the anticommutator is
\beq
L_f=
\{b_0,\cQ_f\}
={1\over 2\pi i}\oint dw 
\left[(T(w)-\partial j_{gh}(w))f(w) + {f'(w)^2\over f(w)}\right]\,,
\eeq
and the gauge fixed kinetic term is just $c_0L_f$.

The combination $T-\partial j_{gh}$ is a modification of the energy 
momentum tensor which corresponds to a change in the central charge 
of the $b,c$ system to $-2$. This is the same twist as applied in 
\cite{Gaiotto:2001ji} to find the sliver solution to the ghost 
equation of motion with the pure ghost action. Using the bosonized 
ghost $\varphi$, with $c\sim \exp i\varphi$ and $b\sim \exp -i\varphi$, this 
amounts to changing the world-sheet action by
\beq
{i\over 2\pi}\int d^2\sigma\sqrt\gamma\,R^{(2)}
\left(\varphi+\bar\varphi\right)\,,
\eeql{correction}
where $R^{(2)}$ is the Ricci scalar on the world-sheet.

This family of kinetic operators is very similar to those proposed in 
\cite{Drukker:2002ct}. There it was argued that the inverse propagator 
should be the convolution of the energy-momentum tensor with an 
arbitrary function $f$, symmetric around the midpoint. If $f$ was 
regular the Feynman graphs would correspond to open surfaces, and 
therefore will describe the theory around the D-brane vacuum. If 
$f$ vanishes at the boundary of the world-sheet the propagator 
would correspond to a world-sheet where the boundary shrunk to a 
point, and should describe purely closed surfaces.\footnote{It is 
not clear what the interpretation is of a zero of $f$ away from 
the boundary $\sigma=0,\pi$.}

More precisely, the propagator
\beq
b_0\int dt \exp\left(-tL_f\right)\,,
\eeq
generates a curved worldsheet with metric 
$ds^2=d\sigma^2+f(\sigma)^2dt^2$. This represents a segment of the 
round sphere if one chooses $f=\sin\sigma$ as in \cite{Drukker:2002ct}, 
or other metrics on it for different functions.

In \cite{Drukker:2002ct} the kinetic operator was constructed from 
the regular energy-momentum tensor, but in fact much of the ghost 
structure was not investigated there closely. Starting from the 
exact gauge invariant $\cQ_f$ we learn that we should use the 
modified energy-momentum tensor, and the extra constant from the 
pure ghost term. All the arguments go through, only that we should 
use the modified world-sheet action on the Riemann surfaces generated 
by the Feynman rules of \cite{Drukker:2002ct}, with the extra constant 
modifying the measure.

There is actually a very nice interpretation of the extra ghost 
piece. One can choose the function $f$ to be almost $1$ everywhere, 
except at the boundaries, where it should vanish. The worldsheet 
generated by $L_f$ will have this profile, that is look like a 
capped cylinder. The constant term $\int (f')^2/f$ will be very large, 
and will favor short propagators. So the worldsheet will be built out 
of very narrow capped cylinders, much like in \cite{Gaiotto:2001ji} 
(though there the cylinders were not capped). It was suggested 
that at the end of those cylinders there should be a vertex operator 
for a zero momentum dilaton.

This world-sheet is flat almost everywhere, except at the ends of the 
cylinders, where one should include the correction term in the action 
(\ref{correction}). That amounts to inserting $c\bar c$ ghosts at the 
end of the cylinder, as is appropriate for a closed string vertex 
operator.

In principle one could now use these Feynman rules and try to derive 
the closed string scattering amplitude on the sphere. We will not 
pursue this here.

\section{Closed string vertices}

In the vacuum of string field theory we expect to find no on-shell 
open string states. But the theory should not be empty, rather it 
should describe closed strings. As explained in 
\cite{Hashimoto:2001sm,Gaiotto:2001ji}, closed strings should be 
the gauge invariant composite operators of the theory, 
and closed string scattering is the correlator of those operators. 
The relevant operators describing closed strings were constructed in 
\cite{Shapiro:ac,Shapiro:gq,Zwiebach:1992bw} as follows. For any 
closed string vertex operator $V=c\bar c V_m$, where $V_m$ is a 
primary of dimension $(1,1)$ in the matter conformal field theory we 
define
\beq
O_V=\int V\left({\pi\over2}\right)\Phi\,.
\eeq
So this is an insertion of a closed string vertex operator at the 
string midpoint.

The crucial point is that those operators should be gauge invariant 
if and only if $V$ describes a physical closed string. So these 
operators allow to add on-shell closed strings to the theory which 
includes off-shell open strings.

The linear gauge transformations around the D-brane vacuum amount to 
a shift of $\Phi$ by a BRST exact field $Q_B\Lambda$. If $Q_B$ 
commutes with $V(\pi/2)$ we can write the variation as
\beq
\delta_\Lambda O_V
=\int Q_B V\left({\pi\over2}\right)\Lambda=0\,,
\eeq
since the integral of a pure gauge quantity vanishes. Therefore the 
mass-shell condition for closed strings should be $[Q_B,V(\pi/2)]=0$, 
as is indeed the case.

This can be used as a check for the kinetic operator of the closed 
string vacuum. 
It should satisfy $[\cQ,V(\pi/2)]=0$ if and only if $V$ is the 
vertex operator for an on-shell closed string. The pure ghost 
operator \cite{Gaiotto:2001ji} commutes with all operators that do 
not involve the anti-ghost, and therefore does not constrain the 
spectrum of closed strings.

All the operators $\cQ_f$ of \cite{Takahashi:2002ez} satisfy this 
condition. The function $f$ has to satisfy $f(\pi/2)=1$ and 
$f'(\pi/2)=0$. So near the midpoint $\cQ_f\sim\int j_B$, which is the 
same as $Q_B$, and therefore the singular terms in the OPE of the 
current $j_Bf$ with any vertex operator $V$ at the midpoint will not 
depend on $f$. We conclude that all those operators satisfy the 
important constraint
\beq
\left[\cQ_f,V\left({\pi\over2}\right)\right]=0
\quad
\Leftrightarrow
\quad
\hbox{$V$ represents an on-shell closed string.}
\eeq

As we noted above, the pure ghost operator is a limit of $\cQ_f$ as $f$ 
approaches a singular function which is zero everywhere but at the 
midpoint. For any smooth approximation of this behavior the condition 
will be satisfied. But in the limit, when we ignore the $j_B$ 
contribution to $\cQ$ and retain only the ghost part, this property 
is lost.

\section{Analytic structure and new star product}

The claim is that $\cQ_f$ is equivalent to the standard BRST operator 
$Q_B$ for a regular $f$, but is different if $f$ has singularities. 
One argument comes from the Feynman rules derived from the gauged fixed 
action, where the condition that $f$ vanish on the boundary shrinks 
the boundaries of the world-sheet to a point. Another explanation was 
that the operator $e^{q(h)}$ used to relate the two operators is ill 
defined.

We want to examine this issue further, and try to see if we can 
still define a similarity operator $e^{q(h)}$ and understand how the 
zeros of $f$ effect the physics so profoundly.

Let us start with a regular function $f$. Then we know that 
$\cQ_f=e^{q(h)}Q_Be^{-q(h)}$. Thus we can write the action as
\ber
S&=&\int \Phi\star e^{q(h)} Q_B e^{-q(h)}\Phi
+{2\over3}\Phi\star\Phi\star\Phi
\nonumber\\*
&=&\bra{V_2}e^{q(h)} Q_B\ket{\Phi'}_1e^{q(h)}\ket{\Phi'}_2
+{2\over 3}\bra{V_3}
e^{q(h)}\ket{\Phi'}_1e^{q(h)}\ket{\Phi'}_2e^{q(h)}\ket{\Phi'}_3\,.
\eer
In the second line we wrote the action in terms of the two and three 
string gluing vertices $\bra{V_2}$ and $\bra{V_3}$ and we defined 
$\Phi'=e^{-q(h)}\Phi$.

Now we can follow \cite{Rastelli:2000iu} and act with the two 
$e^{q(h)}$ to the left on $\bra{V_2}$, and with the three on $\bra{V_3}$. 
Each $q(h)$ is an integral over a current along the 
boundary of the string. These boundaries are glued to each other, and 
apart for a numerical anomaly the sum of all the contours is a trivial 
closed curve.

For example, for the three string vertex we can conformally map the 
three discs describing the string fields to the entire plane (actually 
they are three half discs mapped to the unit disk, but we double the 
string fields to discs, and the final outcome to the entire plane to 
represent the antiholomorphic pieces). The midpoint 
of all the strings is mapped to the origin and each string fills a 
$2\pi/3$ wedge in the plane. The contours defining the three operators 
$q(h)$ are shown by the curves with arrows in \figur{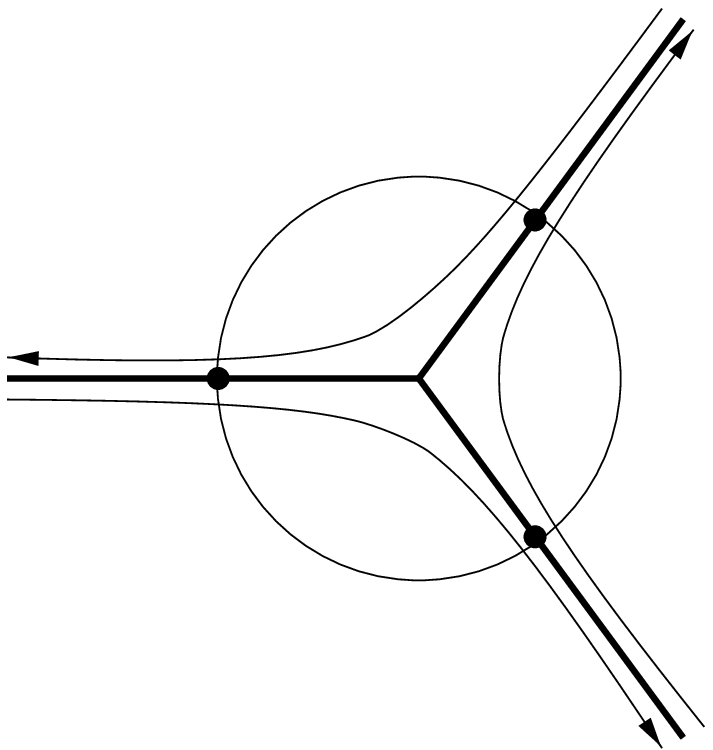}
{The disc is made up of three strings glued along the bold lines. 
The operators $q(h)$ are given by integrals along the curves with the 
arrows and act on each of the strings (the antiholomorphic contribution 
is represented by the image outside the disc). Their sum is zero, 
since the combination of the three contours is trivial.}
The sum of these three curves is clearly trivial.

What goes wrong if the function $f$ is singular on the unit circle? 
By the conformal map the unit circles 
are the curves along which the strings are glued, depicted by the 
thick lines. The function $h=\log f$ will have branch cuts 
going between these line and some point in the interior of the string.

The contours actually pass right through the singularities, so some 
regularization is needed. One option is to move the singularity away 
from the bold line, the other is to move the contours. The first option 
means that we changed $h$, or $f$, so that there is no longer a 
singularity along the unit circle. That will clearly lead to the 
trivial result. So let us try the other regularization, and move 
the contour away from the unit circle, slightly into the three discs. 
This is the picture drawn in \figur{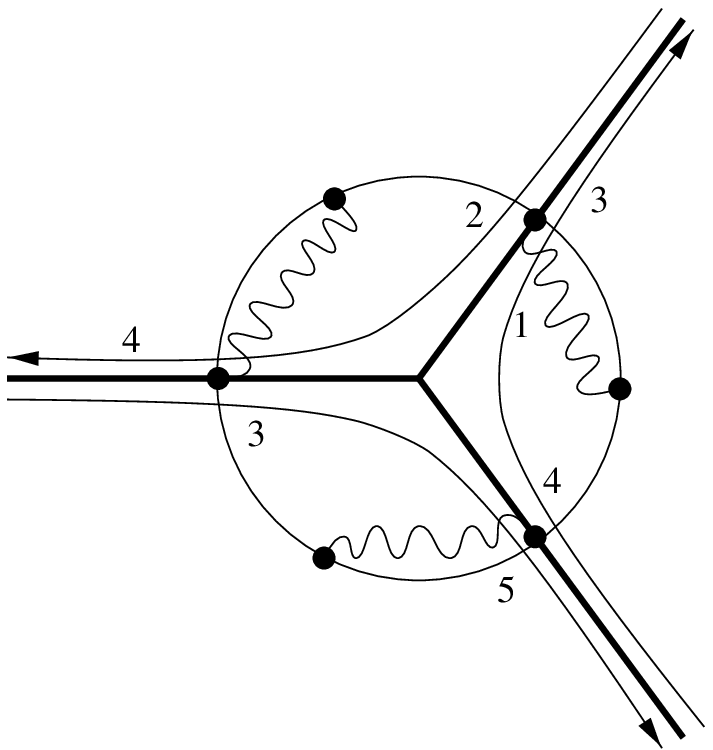}
{If the function $f$ has zeros along the bold lines, where the 
three strings are glued, there will be branch cuts for the function $h$. 
Those are indicated by the wiggly line in the picture. The contour 
defining $q(h)$ passes through 
the branch cuts and the three contours are not trivial anymore. In 
this picture we included numbers labeling a certain choice of sheets 
for the logarithms, assuming $f$ has a double zero.}

With this regularization the contours do not cancel each other, and 
in fact do not close, but end on a different branch of the logarithms. 
Yet we can calculate the total integral, since the difference between 
two sheets of a log is just $2\pi i2n$, where $2n$ is the 
order\footnote{An analytic function satisfying the symmetry properties 
must have the same number of zeros in the upper and lower half plane. 
If the zero is at the boundary, at $w=\pm1$ it must be a double 
zero. The function $|\sin\sigma|$ chosen throughout most of 
\cite{Drukker:2002ct} is not analytic, and probably not a good choice.} 
of the zero of $f$.
So the three contours cancel each other leaving some 
remnant integral of the ghost current times a constant. Since the 
curves do not close, but end on different sheets the result will 
depend on the chosen starting point and sheet assignment. This 
arbitrariness should cancel in the final result, but it is not clear how.

If we choose to start the integration near the midpoint on one of 
the three curves and the sheets depicted by the numbers in Fig.~2. 
we will be left with the net integral shown 
in \figur{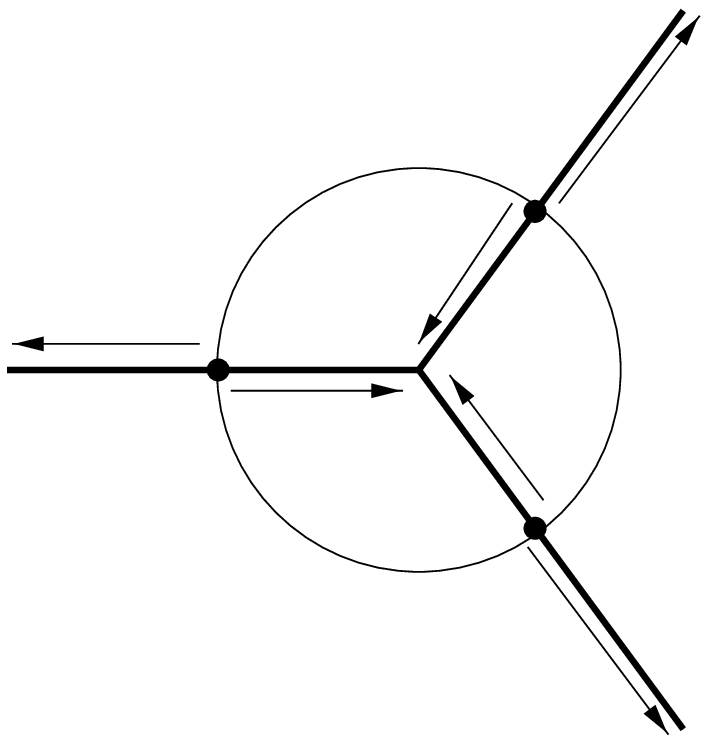}
{With the choice of sheets for the logarithms as in Fig.~2. one is left 
with the integral over the ghost current along the lines indicated 
in this picture. This is the most symmetric choice of 
sheets, but is not unique.}
This amounts to acting on each of the three strings with the operator
\beq
\exp n\left[\int_1^i dw j_{gh}(w)+\int_{-1}^{-i} dw j_{gh}(w)\right]
\eeq
The contours lead from the singular point (the boundary) to the midpoint.

In terms of the bosonized ghost we can write $j_{gh}=i\partial\varphi$, 
so we get only the boundary term
\beq
e^{in\varphi(i)}e^{-in\varphi(1)}e^{in\varphi(-i)}e^{-in\varphi(-1)}\,.
\eeql{integ}
For $n=1$ these are just insertions of $c\bar c$ at the midpoint and 
$b$ and $\bar b$ at the boundaries.

A similar thing happens to the two string vertex $\bra{V_2}$. Again 
one gets those extra $b$ and $c$ terms at the midpoint and boundaries.

It is natural to absorb these ghost insertions in the definition of 
the two and three string vertices, or the star product. Thus we learn 
that even for a singular $f$ we can replace $\cQ_f$ with the old $Q_B$, 
but the price is that we have to modify the star product (and integration). 
The new star product identifies the half strings as before, but also 
inserts $b$ and $c$ ghosts at the boundary and midpoints.

So we found a new form for the action of string field theory around the 
vacuum, where instead of modifying the kinetic operator, we changed 
the gluing rules, by adding extra ghosts. As mentioned above, there is 
some arbitrariness in choosing the starting point in the contour 
integrals of $q(h)$, which will change the type of ghost insertions. 
We expect this to be some symmetry that will not show up in any 
calculation, but the mechanism that cancels this is not clear.

One other simple prescription is to start the integral at one of the 
boundaries. Then all the integrals as in Fig. 3. and equation 
(\ref{integ}) will run between 
the boundary points, and not end at the midpoint. For the two string 
vertex it amounts to inserting $b\partial b$ at one boundary and 
$c\partial c$ at the other. In the three point vertex two boundary 
points get $b\partial b$ each and the third boundary point will get 
$c\partial c\partial^2 c\partial^3 c$. This is a less symmetric 
prescription, but involves fewer ghost insertions, so it might be easier 
to work with.

The upshot of this approach is that the dependence on the form of $f$ 
completely disappeared. The only remnant is the singularity structure, 
or the order of the zero (or more generally zeros) of $f$. So clearly 
all functions $f$ with a double zero at the boundary, like 
$f=-(w-1/w)^2/4$ studied in \cite{Takahashi:2002ez,Kishimoto:2002xi}
are equivalent to each other. That is, they are related by small 
gauge transformations.

It is natural to conjecture that more zeros, or higher order zeros 
correspond to higher rank projectors, and those should give multiple 
D-branes. Thus we can write string field theory around any classical 
solution in terms of the usual $Q_B$, but with different gluing rules.

\section{Discussion}

In this paper we studied the kinetic operators $\cQ_f$ for the vacuum 
of string field theory constructed by Takahashi and Tanimoto 
\cite{Takahashi:2002ez}. They derived them from a shift by a formal 
classical solution of open cubic string field theory, 
and in \cite{Kishimoto:2002xi} it was shown that for a certain 
function $f$ the cohomology is trivial (at ghost number one).

These operators can serve instead of the pure ghost operator of 
\cite{Gaiotto:2001ji}. There is a large family of operators that 
are analogous to each other and are less singular than the pure ghost 
midpoint insertion. But we also show that the pure ghost operator 
is a limit of $\cQ_f$ for a singular function $f$.

We looked further into those operators. First we explained the origin 
of the singularity in the field redefinition that relates those 
$\cQ_f$ to the BRST operator $Q_B$. Those two operators are gauge 
equivalent if the function $f$ is regular on the unit circle, but 
if $f$ has a zero the similarity transformation is singular.

Another check on these kinetic operators are the propagators derived 
from them. By choosing the Siegel gauge one reduces to a gauge fixed 
action of the form proposed in \cite{Drukker:2002ct} from very 
different considerations. The fact that the function $f$ vanishes 
causes the boundaries of the world-sheet to shrink to points. So 
the Feynman graphs of the theory will be surfaces without 
boundaries, as is appropriate for closed strings.

The Feynman rules derived from $\cQ_f$ give a very explicit procedure 
for calculating the closed string scattering amplitudes. The geometries 
of the world-sheets were already found in \cite{Drukker:2002ct} and the 
ghost part of the world-sheet action and measure, which were not 
studied there, are given here. One could use these rules to calculate 
closed string scattering as was done for open strings in 
\cite{Giddings:1986iy}.

We proceeded to look at the singular field redefinition that relates 
$Q_B$ to $\cQ_f$. We propose that even when $f$ is singular one can 
replace one operator by another, at the price of changing the gluing 
rules that define the star product and integration. This leads to 
a new form of the action with extra ghost inserted when string fields 
are glued.

This formalism deserves further study, as we plan to do, but one can 
already see some nice features. One is that the only information about 
$f$ that enters into the new action is the singularity structure, 
i.e. the number and locations of zeros and poles. All functions $f$ 
with the same singularities should describe the same vacuum of the 
theory. The new formulation we propose makes this fact manifest.

This transformation is similar in some ways to the Seiberg-Witten map 
of non-com\-mu\-tative gauge theory \cite{Seiberg:1999vs}. That is 
the equivalence of two theories, one with some background field turned 
on (effecting the kinetic term) with another theory with no background 
field, but a different multiplication rule. Here too we can formulate 
the theory with either a new kinetic operator, or a new multiplication 
rule.

For a certain function $f(\sigma)=\sin^2\sigma$, there is no cohomology 
at ghost number one. Therefore a double zero at the boundary should 
correspond to the closed string vacuum. We conjecture that a different 
number (or order) of zeros of $f$ will correspond to different vacua of 
the theory, and in particular other functions will correspond to 
vacua with more D-branes.

If we relate the order of the zeros of $f$ to the number of D-brane, 
it is tempting to think that there should be a way to calculate the 
action of the brane from it. Perhaps using the techniques similar to 
those of Section~5 one could express the action for the classical 
solution in terms of an index of the function $f$. So far we have 
been unable to do that.

\section*{Acknowledgments}

I would like to thank Ofer Aharony, Micha Berkooz, 
Sunny Itzhaki, Leonardo Rastelli, Tomohiko Takahashi and Barton Zwiebach 
for very useful discussions.


\end{document}